\documentclass[preprintnumbers,eqsecnum,aps,prd,epsf,showpacs,nofootinbib]{revtex4}
\usepackage{latexsym,amsmath,amssymb}
\usepackage[dvips]{graphicx,color}
\usepackage{dcolumn,bm}    
\usepackage{subfigure}  

\newcommand{\ma}[1]{{\mathrm{#1}}}

\newcommand{\calO}{{\cal O}}
\newcommand{\calJ}{{\cal J}}
\newcommand{\cale}{{\cal E}}
\newcommand{\calA}{{\cal A}}
\newcommand{\pa}{{\partial}}
\newcommand{\na}{{\nabla}}
\newcommand{\calg}{{\cal G}}
\newcommand{\frR}{{\frak R}}

\begin{document}
\thispagestyle{empty}
\title{Nonlinear superhorizon curvature perturbation 
in generic single-field inflation}

\preprint{YITP-13-1, RUP-12-12}

\author{Yu-ichi Takamizu$^{1}$}
\email{takamizu_at_yukawa.kyoto-u.ac.jp}

\author{Tsutomu Kobayashi$^{2}$}
\email{tsutomu_at_rikkyo.ac.jp}
\affiliation{
\\
$^{1}$ Yukawa Institute for Theoretical Physics
 Kyoto University, Kyoto 606-8502, Japan
\\
$^{2}$ Department of Physics, Rikkyo University, Toshima, Tokyo 175-8501, Japan
}
\date{\today}

\begin{abstract}
 We develop a theory of nonlinear cosmological perturbations on
 superhorizon scales for generic single-field inflation.
Our inflaton is described by the Lagrangian of the form 
 $W(X,\phi)-G(X,\phi)\Box\phi$ with
$X=-\partial^{\mu}\phi\partial_{\mu}\phi/2$,
which is no longer equivalent to a perfect fluid.
This model is more general than k-inflation,
and is called G-inflation.
 A general nonlinear solution for the metric and the scalar field is obtained at
 second order in gradient expansion.
 We derive a simple master equation governing the large-scale evolution of
 the nonlinear curvature perturbation.
 It turns out that the nonlinear evolution equation
 is deduced as a straightforward extension of the corresponding
 linear equation for the curvature perturbation on uniform $\phi$ 
hypersurfaces. 
\end{abstract}
\maketitle

\section{Introduction}
\label{sec:intro}

Non-Gaussianity in primordial fluctuations is
one of the most powerful tools to
distinguish different models of inflation
(see, {\em e.g.}, Ref.~\cite{CQG-focus-NG} and references therein).
To quantify the amount of non-Gaussianity and clarify non-Gaussian
observational signatures, it is important to develop
methods to
deal with nonlinear cosmological perturbations. 
Second-order perturbation 
theory~\cite{Maldacena:2002vr,Acquaviva:2002ud,Malik:2003mv}
is frequently used for this purpose. 
Spatial gradient expansion is employed as well in the literature~\cite{Lifshitz:1963ps,
Starobinsky:1986fxa,Salopek:1990jq,Deruelle:1994iz,Nambu:1994hu,
Sasaki:1995aw,Sasaki:1998ug, 
Kodama:1997qw, Wands:2000dp, Lyth:2004gb, 
Lyth:2005du,
Rigopoulos:2003ak,Rigopoulos:2004gr,Lyth:2005fi,Takamizu:2010je,Takamizu:2010xy,Takamizu:2008ra,Tanaka:2007gh,Tanaka:2006zp,Naruko:2012fe,Sugiyama:2012tj,Narukosasaki,Gao}. 
The former can be used to describe the generation and evolution
of primordial perturbations inside the horizon,
while the latter can deal with the classical {\em nonlinear} evolution after horizon exit. Therefore, it is important to develop both methods and to use them
complementarily.

In this paper, we are interested in the classical evolution of
nonlinear cosmological perturbations on superhorizon scales.
This has been addressed extensively in the context of
the separate universe approach~\cite{Wands:2000dp} or, equivalently, 
the $\delta N$ formalism~\cite{Starobinsky:1986fxa,Sasaki:1995aw, 
Sasaki:1998ug, Lyth:2005fi}. 
The $\delta N$ formalism is the zeroth-order truncation of gradient expansion. 
However, higher-order contributions in gradient expansion
can be important for extracting more detailed information about
non-Gaussianity from primordial fluctuations.
Indeed, it has been argued that,
in the presence of a large effect from the decaying mode due to slow-roll violation,
the second-order corrections
play a crucial role in the superhorizon evolution of the curvature perturbation
already at linear order~\cite{Seto:1999jc, Leach:2001zf,Jain:2007au},
as well as at nonlinear order~\cite{Takamizu:2010xy,Takamizu:2010je}. 
In order to study time evolution of the 
curvature perturbation on superhorizon 
scales in the context of non-Gaussianity, 
it is necessary to develop nonlinear theory of 
cosmological perturbations valid up to second order in spatial 
gradient expansion.

The gradient expansion technique
has been applied up to second order
to a universe dominated by a canonical scalar field and by a perfect fluid with
a constant equation of state parameter, $P/\rho=$const~\cite{Tanaka:2006zp,Tanaka:2007gh}.
The formulae have been extended to be capable of
a universe filled with a non-canonical scalar field described by a generic
Lagrangian of the form $W(-\partial^{\mu}\phi\partial_{\mu}\phi/2,\phi)$,
as well as a universe dominated by a perfect
fluid with a general equation of state, $P=P(\rho)$~\cite{Takamizu:2008ra,Takamizu:2010xy}. Those systems are characterized by a single scalar degree of freedom, and hence one expects that a single master variable governs
the evolution of scalar perturbations even at nonlinear order.
By virtue of gradient expansion, one can indeed derive a simple evolution equation
for an appropriately defined master variable $\frR_u^{\rm NL}$:
\begin{eqnarray}
{{\frR}_u^{\rm NL}}''+2 {z'\over z} 
{{\frR}_u^{\rm NL}}' +{c_s^2\over 4} \,{}^{(2)}R[\,
{\frR}_u^{\rm NL}\,]=\calO(\epsilon^4)\,,\label{ieq1}
\end{eqnarray}
where the prime represents differentiation with respect to the conformal time,
$\epsilon$ is the small expansion parameter, and the other quantities will be
defined in the rest of the paper. This equation is to be compared with its linear counterpart:
\begin{eqnarray}
{{\cal R}^{\rm Lin}_u}''+2{z'\over z} {{\cal R}^{\rm Lin}_u}'
-c_s^2\,\Delta {\cal R}^{\rm Lin}_u=0,\label{ieq2}
\end{eqnarray}
from which one notices
the correspondence between the linear and nonlinear evolution equations. 
Gradient expansion can be applied also to
a multi-component system, yielding the formalism ``beyond $\delta N$''
developed in a recent paper~\cite{Naruko:2012fe}.

The purpose of this paper is to extend the gradient expansion formalism further
to include a more general class of scalar-field theories
obeying a second-order equation of motion. 
The scalar-field Lagrangian we consider is of the form
$W(-\partial^{\mu}\phi\partial_{\mu}\phi/2,\phi)-G(-\partial^{\mu}\phi\partial_{\mu}\phi/2,\phi)\Box\phi$.
Inflation driven by this scalar field is more general than k-inflation
and is called G-inflation.
It is known that
k-inflation \cite{ArmendarizPicon:1999rj, Garriga:1999vw} with the Lagrangian $W(-\partial^{\mu}\phi\partial_{\mu}\phi/2,\phi)$ is
equivalently described by a perfect fluid cosmology.
However, in the presence of $G\Box\phi$,
the scalar field is no longer equivalent to a perfect fluid
and behaves as a {\em imperfect} fluid~\cite{KGB,KGB2}. The authors of Ref.~\cite{Narukosasaki} investigated superhorizon conservation of
the curvature perturbation from G-inflation at zeroth order in gradient expansion,
and Gao worked out the zeroth-order analysis in the context of
the most general single-field inflation model~\cite{Gao}.
In this paper, we present a general analytic solution
for the metric and the scalar field for G-inflation at second order in gradient expansion.
By doing so we
extend the previous result for a perfect fluid \cite{Takamizu:2008ra} and
show that
the nonlinear evolution equation of the form~(\ref{ieq1})
is deduced straightforwardly from the corresponding linear result even in
the case of G-inflation.

This paper is organized as follows.
In the next section, we define the non-canonical scalar-field theory that we consider
in this paper.
In Sec.~\ref{sec:formulation}, we develop a theory of nonlinear 
cosmological perturbations on superhorizon scales
and derive the field equations employing gradient expansion. 
We then integrate the relevant field equations
to obtain a general solution for the metric and the scalar field
in Sec.~\ref{sec:solution}.
The issue of defining appropriately the nonlinear curvature perturbation is addressed
and the evolution equation for that variable is derived
in Sec.~\ref{sec:nlcurvpert}. 
Section~\ref{sec:summary} is devoted to a
summary of this paper and discussion. 

\section{G-inflation}
\label{sec:scalarfield}

In this paper we study a generic inflation model driven by a single scalar field.
We go beyond k-inflation for which the Lagrangian for the scalar field is
given by an arbitrary function of $\phi$ and $X:=-g^{\mu\nu}\partial_\mu\phi\partial_\nu\phi/2$,
${\cal L}=W(X, \phi)$,
and consider the scalar field described by
%
\begin{equation}
 I = \int d^4x\sqrt{-g}\bigl[ W(X,\phi)-G(X,\phi)\Box\phi\bigr], \label{scalar-lag}
\end{equation}
where $G$ is also an arbitrary function of $\phi$ and $X$.
We assume that $\phi$ is minimally coupled to gravity.
Although the above action depends upon the second derivative of $\phi$ through
$\Box\phi =g^{\mu\nu}\na_\mu\na_\nu\phi$,
the resulting field equations for $\phi$ and $g_{\mu\nu}$ remain second order.
In this sense the above action gives rise to a more general single-field inflation model
than k-inflation, {\em i.e.}, G-inflation~\cite{Kobayashi:2010cm,Kobayashi:2011nu,Kobayashi:2011pc}. The same scalar-field Lagrangian is used
in the context of dark energy and called kinetic gravity braiding~\cite{KGB}. 
Interesting cosmological applications of the Lagrangian~(\ref{scalar-lag})
can also be found, {\em e.g.}, in~\cite{Mizuno:2010ag, Kimura:2010di, Kamada:2010qe,
GBounce, Kimura:2011td, Qiu:2011cy, Cai:2012va, Ohashi:2012wf}.
In fact, the most general inflation model with second-order field equations
was proposed in~\cite{Kobayashi:2011nu} based on Horndeski's scalar-tensor
theory~\cite{Horndeski, GenG}.
However, in this paper we focus on
the action~(\ref{scalar-lag}) which
belongs to a subclass of the most general single-field inflation model,
because it involves sufficiently new and interesting ingredients
while avoiding unwanted complexity.
Throughout the paper we use Planck units, $M_{\rm pl}=1$,
and assume that the vector $-g^{\mu\nu}\na_{\nu}\phi$
is timelike and future-directed.
(The assumption is reasonable because we are not interested in
a too inhomogeneous universe.)

The equation of motion for $\phi$ is given by 
%
\begin{equation}
\nabla_{\mu}\Bigl[(W_X-G_\phi-G_X \Box\phi)\na^\mu \phi-G_X 
\na^\mu X\Bigr]+W_\phi-G_\phi\Box\phi= 0,
\end{equation}
where the subscripts $X$ and $\phi$ stand for differentiation with respect to
$X$ and $\phi$, respectively.
More explicitly, we have
\begin{eqnarray}
&&W_X\Box\phi-W_{XX}(\na_\mu\na_\nu \phi)(\na^\mu\phi\na^\nu \phi)-
2W_{\phi X}X+W_\phi-2(G_\phi-G_{\phi X}X)\Box\phi\nonumber\\
&&+G_X\left[(\na_\mu\na_\nu \phi)(\na^\mu\na^\nu \phi)-(\Box\phi)^2+R_{\mu\nu}\na^\mu\phi\na^\nu\phi\right]+2G_{\phi X}(\na_\mu\na_\nu \phi)(\na^\mu\phi\na^\nu \phi)+2G_{\phi\phi}X\notag\\
&&-G_{XX}(\na^\mu\na^\lambda\phi-g^{\mu\lambda}\Box\phi)(\na_\mu\na^\nu\phi)\na_\nu\phi\na_\lambda\phi=0.
  \label{eqn:EOM-phi}
\end{eqnarray}
The energy-momentum tensor of the scalar
field is given by 
%
\begin{equation}
 T_{\mu\nu} = W_X\na_\mu\phi\na_\nu\phi+Wg_{\mu\nu}-(\na_\mu G \na_\nu \phi
+\na_\nu G \na_\mu \phi) +g_{\mu\nu} \na_\lambda G\na^\lambda \phi-
G_X\Box\phi\na_\mu\phi\na_\nu\phi.
  \label{eqn:Tmunu-phi}
\end{equation}
It is well known that k-inflation
allows for an equivalent description in terms of a perfect fluid, {\em i.e.},
the energy-momentum tensor reduces to that of a perfect fluid with a
four-velocity $u_\mu=\partial_\mu\phi/\sqrt{2X}$.
However, as emphasized in~\cite{KGB},
for $G_X\neq 0$ the energy-momentum tensor cannot be expressed
in a perfect-fluid form in general.
This {\em imperfect} nature
characterizes the crucial difference between G- and k-inflation.


\section{Nonlinear cosmological perturbations}
\label{sec:formulation}

In this section we shall develop a theory of nonlinear cosmological
perturbations on superhorizon scales, following Ref.~\cite{Takamizu:2008ra}.
For this purpose we employ the
Arnowitt-Deser-Misner (ADM)
formalism and perform a gradient expansion 
in the uniform expansion
slicing and the time-slice-orthogonal threading.

\subsection{The ADM decomposition}

Employing the ($3+1$)-decomposition of the metric, we write
%
\begin{equation}
 ds^2 = g_{\mu\nu}dx^{\mu}dx^{\nu}
  = - \alpha^2 dt^2 + \hat{\gamma}_{ij}(dx^i+\beta^idt)(dx^j+\beta^jdt), 
\end{equation}
where $\alpha$ is the lapse function and $\beta^i$ is the shift vector.
Here, latin indices run over $1, 2, 3$. We introduce 
the unit vector $n^{\mu}$ normal to the constant $t$ hypersurfaces,
%
\begin{equation}
 n_{\mu}dx^{\mu} = -\alpha dt, \quad
  n^{\mu}\partial_{\mu} = \frac{1}{\alpha}(\partial_t-\beta^i\partial_i).
\end{equation}
The extrinsic curvature $K_{ij}$ of constant $t$ hypersurfaces is given by
%
\begin{equation}
 K_{ij} =\na_i n_j=
\frac{1}{2\alpha}\left(\partial_t\hat{\gamma}_{ij}-\hat{D}_i\beta_j-
\hat{D}_j\beta_i\right),
  \label{eqn:def-K}
\end{equation}
where $\hat{D}_i$ is the covariant derivative associated with
the spatial metric $\hat{\gamma}_{ij}$. 
The spatial metric and the extrinsic curvature can further be
expressed in a convenient form as
\begin{align}
 \hat{\gamma}_{i j} &= a^2(t) e^{2\psi(t, \mathbf{x})} \gamma_{i j} ,
\\
 K_{i j} &= a^2(t) e^{2\psi} \left( \frac{1}{3} K \gamma_{i j}
 + A_{i j} \right),
\label{deco-Kij}
\end{align}
where $a(t)$ is the scale factor of a fiducial homogeneous 
Friedmann-Lema\^itre-Robertson-Walker (FLRW) spacetime,
the determinant of $\gamma_{ij}$ is normalised to unity, $\ma{det}\,\gamma_{ij}=1$,
and $A_{i j}$ is the trace-free part of $K_{ij}$, $\gamma^{ij}A_{ij}=0$. 
The trace $K:= \hat\gamma^{ij}K_{ij}$ is explicitly written as
\begin{gather}
 K
 =  \frac{1}{\alpha} \Bigl[ 3( H + \pa_t \psi) - \hat{D}_i  
\beta^i \Bigr] \,,  
\label{def-K}
\end{gather}
where $H=H(t)$ is the Hubble parameter defined by
 $H := d \ln a (t)/d t$.
In deriving Eq.~(\ref{def-K})
$\partial_t(\ma{det}\,\gamma_{ij})=\gamma^{ij}\partial_t\gamma_{ij}=0$ was used.
Hereafter, 
in order to simplify the equations we
choose the spatial coordinates appropriately to set 
\begin{align}
\beta^i=0.
\label{time-orthogonal-cond}
\end{align}
We call this choice of spatial coordinates as the {\it time-slice-orthogonal threading}.

With $\beta_i=0$ all the independent components of the
energy-momentum tensor (\ref{eqn:Tmunu-phi}) are expressed as 
%
\begin{eqnarray}
 E & := & T_{\mu\nu}n^{\mu}n^{\nu} = (W_X-G_X\Box\phi)(\pa_\perp\phi)^2-W-
\pa_\perp G\pa_\perp\phi-\hat{\gamma}^{ij}\pa_i G\pa_j \phi, 
\label{def:EEE}\\
 -J_i & := & T_{\mu i}n^{\mu} = \left[(W_X-G_X\Box\phi)\pa_i\phi-\pa_i G\right]\pa_\perp\phi-\pa_\perp G\pa_i \phi, 
\label{def:J}\\
 S_{ij} & := & T_{ij}, 
\end{eqnarray}
where $\partial_{\perp}:= n^{\mu}\partial_{\mu}$.

Let us now move on to the $(3+1)$-decomposition of the Einstein equations.
In the ADM Language, the 
Einstein equations are separated into four constraints
(the Hamiltonian constraint and three momentum constraints)
and six dynamical equations for the spatial metric. The constraints are
%
\begin{align}
& {1\over a^2} R[e^{2\psi}\gamma] + {2\over 3}K^2 - A_{ij}A^{ij} = 2E, 
   \label{eqn:Hamiltonian-const}\\
& {2\over 3}\partial_i K -e^{-3\psi} D_j\left(e^{3\psi} A^j_{\ i}\right) = J_i,
  \label{eqn:Momentum-const}
\end{align}
where $R[e^{2\psi}\gamma]$ is the Ricci scalar constructed from
the metric $e^{2\psi}\gamma_{ij}$ and 
$D_i$ is the covariant derivative with respect to $\gamma_{ij}$.
The spatial indices here are raised or lowered by $\gamma^{ij}$ and $\gamma_{ij}$, 
respectively. 
As for the dynamical equations, 
the following first-order equations for the spatial 
metric ($\psi$, $\gamma_{ij}$) are
deduced from the definition of the extrinsic curvature~(\ref{eqn:def-K}):
%
\begin{eqnarray}
 \partial_{\perp}\psi& = & 
  -\frac{H}{\alpha}+{K \over 3},
  \label{eqn:dpsi}\\
 \partial_{\perp}{\gamma}_{ij} & = & 2{A}_{ij}.
  \label{eqn:dgamma}
\end{eqnarray}
The dynamical equations for the extrinsic curvature ($K$,
${A}_{ij}$) are given by
%
\begin{eqnarray}
 \partial_{\perp}K & = &  
  -\frac{K^2}{3} - {A}^{ij}{A}_{ij} + \frac{\hat{D}^2\alpha}{\alpha} 
  - \frac{1}{2}\left(E+3P\right), 
  \label{eqn:dK}\\
  \partial_{\perp}{A}_{ij} & = & 
   -K{A}_{ij}+2{A}_i^{\ k}{A}_{kj}
+ \frac{1}{\alpha}\left[\hat{D}_i \hat{D}_j \alpha\right]^{{\rm TF}}
   -\frac{1}{a^2e^{2\psi}} \left[R_{ij}[e^{2\psi}\gamma]- S_{ij}\right]^{{\rm TF}},
        \label{eqn:dA}
\end{eqnarray}
where 
\begin{align}
P:= {1\over 3}a^{-2} e^{-2\psi} \gamma^{ij} S_{ij},
\label{def:P-Sii}
\end{align}
$\hat{D}^2:=\hat{\gamma}^{ij}\hat{D}_i\hat{D}_j$, and
$R_{ij}[e^{2\psi}\gamma]$ is the
Ricci tensor constructed from the metric $e^{2\psi}\gamma_{ij}$.
The trace-free projection operator $[\ldots]^{{\rm TF}}$ is defined for
an arbitrary tensor $Q_{ij}$ as 
\begin{align}
\left[Q_{ij}\right]^{{\rm TF}}:= Q_{ij}-{1\over 3} \gamma_{ij}\gamma^{kl}Q_{kl}.
\end{align}

For the purpose of solving the Einstein equations,
the most convenient choice of the temporal coordinate is
such that the expansion $K$ is uniform and takes the form:
%
\begin{equation}
 K(t, \mathbf{x}) = 3H(t). 
 \label{eqn:uniform-Hubble}
\end{equation}
Hereafter we call this gauge choice with (\ref{time-orthogonal-cond}) 
{\em the uniform expansion gauge}. 
Adopting this gauge choice,
Eq.~(\ref{eqn:dpsi}) reduces simply to
\begin{equation}
  \partial_t \psi=H(\alpha-1)=:H\, \delta \alpha(t, \mathbf{x}). 
\label{eqn:uniform-H}
\end{equation}
From this,  if we take the uniform expansion gauge, we can see that  
{\it the time evolution of the curvature perturbation $\psi$ is
caused by the inhomogeneous part of the lapse function
$\delta \alpha(t,x^i)$ only.} It is related to the non-adiabatic perturbation.

\subsection{Gradient expansion: basic assumptions and the order of various terms} 

In the gradient expansion approach, we introduce a flat FLRW universe characterized by
($a(t)$, $\phi_0(t)$) as a background and suppose that the
characteristic length scale $L=a/k$, where $k$ is 
a wavenumber of a perturbation, is longer than the 
Hubble length scale $1/H$ of the background, $HL\gg 1$.
We use $\epsilon:= 1/(HL)=k/(aH)$ as a small parameter
to keep track of the order of various terms and
expand the equations in terms of $\epsilon$, so that
a spatial derivative acting on a perturbation raises the order by $\epsilon$.

The background flat FLRW universe characterized by ($a(t)$, $\phi_0(t)$) satisfies the
Einstein equations,
%
\begin{equation}
 H^2(t) = \frac{1}{3}\rho_0, \quad \dot{H}(t)=-{1\over 2}(\rho_0+P_0),
 \label{eqn:background-H}
\end{equation}
and the scalar-field equation of motion,
\begin{equation}
 \dot{\cal J}_0+3H\calJ_0 =\left(W_\phi-2XG_{\phi\phi}-2G_{\phi X}X\ddot{\phi}\right)_0.
 \label{eqn:background-phi-J}
\end{equation}
Here,
an overdot $(\dot{\ })$ denotes differentiation with respect to $t$, and
a subscript $0$ indicates
the corresponding background quantity, 
{\em i.e.},
$W_0:= W(X_0,\phi_0)$, $(W_{X})_0:= W_X(X_0,\phi_0)$, etc., where $X_0:=\dot{\phi}^2_0/2$. 
The background energy density and pressure, $\rho_0$ and $P_0$,
are given by
\begin{eqnarray}
&&\rho_0=\left[-W+2X(W_{X}+3HG_{X} \dot{\phi}-G_{\phi})\right]_0,\\
&&P_0=\left[W-2X(G_{X}\ddot{\phi}+G_{\phi})\right]_0,
\end{eqnarray}
while
$\calJ_0$
is defined as
\begin{align}
&\calJ_0=\left[W_{X}\dot{\phi}-2G_{\phi}\dot{\phi}+6H G_{X} X\right]_0.
\end{align}
If $W$ and $G$ do not depend on $\phi$, the right hand side of Eq.~(\ref{eqn:background-phi-J})
vanishes and hence ${\cal J}_0$ is conserved. In this case, ${\cal J}_0$
is the Noether current associated with the shift symmetry $\phi\to\phi+c$.
Note that the above quantities may be written in a different way as
\begin{align}
&\rho_0=\calJ_0\dot{\phi}_0-W_0+2(X G_{\phi})_0,\\
&\rho_0+P_0=\calJ_0\dot{\phi}_0-2(G_{X}X\ddot{\phi})_0.
\label{eqn:rho0+P0}
\end{align}
Note also that
the scalar-field equation of motion~(\ref{eqn:background-phi-J}) 
can be written as
\begin{align}
{\cal{G}}\ddot{\phi}_0+3\Theta \calJ_0 +{\cal E}_\phi=0,
\label{eqn:background-gphi}
\end{align}
where 
\begin{align}
&{\cal{G}}(t) := \cale_X-3\Theta (G_{X}\dot{\phi})_0,\\
&\Theta(t)  := H -(G_X X \dot{\phi})_0,\\
&\cale_\phi(t) := \left[2 X W_{X\phi} -W_\phi+
6 H G_{\phi X}X \dot{\phi}- 2X G_{\phi\phi}\right]_0,
\label{def:cale-phi}\\
&\cale_X (t) := \left[W_X+2 XW_{XX}+9H G_X \dot{\phi}+6 H G_{XX} 
\dot{\phi}_0 -2 G_\phi-2 X G_{\phi X}\right]_0.\label{def:cale-X}
\end{align}
These functions will also be used later.

Since the background
FLRW universe must be recovered
at zeroth order in gradient expansion,
the spatial metric must take the locally homogeneous and isotropic form
in the limit $\epsilon\to 0$.
This leads to the 
following assumption: 
%
\begin{equation}
 \partial_t{\gamma}_{ij} = \calO(\epsilon^2). 
 \label{eqn:assumption-gamma}
\end{equation}
This assumption is justified as
follows~\cite{Lyth:2004gb,Tanaka:2006zp,Tanaka:2007gh,Takamizu:2008ra,Takamizu:2010xy,Takamizu:2010je,Naruko:2012fe,Sugiyama:2012tj}.
If $\partial_t\gamma_{ij}$ were ${\cal O}(\epsilon)$,
the leading term would correspond to homogeneous and anisotropic perturbations,
which are known to decay quickly. We may therefore reasonably
assume $\partial_t\gamma_{ij}={\cal O}(\epsilon^2)$ and not $\partial_t\gamma_{ij}={\cal O}(\epsilon)$.
However, $\psi$ and ${\gamma}_{ij}$
(without any derivatives acting on them) are of order $\calO(1)$.

Using the assumption (\ref{eqn:assumption-gamma}) made above and
the basic equations derived in the previous subsection,
one can now deduce the order of various terms in gradient expansion.
First, from Eq.~(\ref{eqn:dgamma}) we see that 
%
\begin{equation}
 {A}_{ij}=\calO(\epsilon^2). \label{eqn:A-e2}
\end{equation}
Substituting Eq.~(\ref{eqn:A-e2}) into 
Eq.~(\ref{eqn:Momentum-const})
under the gauge condition~(\ref{eqn:uniform-Hubble}), 
we obtain $J_i=\calO(\epsilon^3)$.
Then, this condition combined with the definition~(\ref{def:J})
implies that
$\partial_i\delta\phi=\calO(\epsilon^3)$, where $\delta\phi(t,\mathbf{x})
: = \phi(t,\mathbf{x})-\phi_0(t)$. 
The same equations also imply that $\pa_i G=\calO(\epsilon^3)$. 
By absorbing a homogeneous part of $\delta \phi$ into $\phi_0$ (and redefining
$a(t)$ accordingly), we have 
%
\begin{equation}
 \delta \phi = \calO(\epsilon^2). 
\end{equation}
It is clear from the condition~(\ref{eqn:A-e2}) and
the Hamiltonian constraint~(\ref{eqn:Hamiltonian-const}) that
\begin{align}
\delta E : = E(t,x^i)-\rho_0(t)=\calO(\epsilon^2).
\end{align}
Since the definition~(\ref{def:EEE}) tells that
$E-\rho_0={\rm max}\{\calO(\delta\phi),\, \calO(\pa_\perp \phi-\pa_t\phi_0)\}$,
we see
$\partial_t(\delta\phi)-\dot{\phi}_0 \delta\alpha =\calO(\epsilon^2)$, leading to
%
\begin{equation}
 \delta \alpha=\calO(\epsilon^2).
\end{equation}
Then, it follows immediately from Eq.~(\ref{eqn:uniform-H}) that
\begin{equation}
 \pa_t \psi = \calO(\epsilon^2). 
\end{equation}
Similarly, for the spatial energy-momentum component
we find
\begin{equation}
 \delta P : = P(t,\mathbf{x})-P_0(t) = 
\calO(\epsilon^2). 
\end{equation}

In summary, we have evaluated the order of various quantities as follows:
%
\begin{eqnarray}
 & & 
  \psi = \calO(1), \quad {\gamma}_{ij} = \calO(1), 
\nonumber\\
 & & 
  \delta \alpha = \calO(\epsilon^2),  \quad \delta \phi=\calO(\epsilon^2),  
  \quad \delta E=\calO(\epsilon^2), 
  \quad \delta P = \calO(\epsilon^2), \quad {A}_{ij} = \calO(\epsilon^2),\nonumber\\
 & & 
  \partial_t{\gamma}_{ij}=\calO(\epsilon^2), \quad
  \partial_t\psi=\calO(\epsilon^2), \quad
  \beta^i = \calO(\epsilon^3),  \quad \pa_i G = \calO(\epsilon^3), \quad [S_{ij}]^{{\rm TF}} = \calO(\epsilon^6),
  \label{eqn:order-of-magnitude}
\end{eqnarray}
where the assumptions made have also been included.

\subsection{Field equations up to ${\cal O}(\epsilon^2)$ in gradient expansion}

Keeping the order of various terms~(\ref{eqn:order-of-magnitude}) in mind,
let us derive the governing equations in the uniform expansion gauge.
%
The Hamiltonian and momentum constraints are
%
\begin{eqnarray}
 R[e^{2\psi}\gamma]& = & 2\delta E + \calO(\epsilon^4),
  \label{eqn:Hamiltonian}\\
 e^{-3\psi}{D}_j\left(e^{3\psi}{A}_{i}^{\ j}\right)
  & = & -J_i + \calO(\epsilon^5).
  \label{eqn:Momentum}
\end{eqnarray}
The evolution equations for the spatial metric are given by
%
\begin{eqnarray}
 {\partial_t\psi}  =  H\delta \alpha + \calO(\epsilon^4), \qquad 
\partial_t{\gamma}_{ij}  =  2{A}_{ij}+\calO(\epsilon^4),
  \label{eqn:evol-gamma}
\end{eqnarray}
while the evolution equations for the extrinsic curvature are 
%
\begin{eqnarray}
 \partial_t{A}_{ij}& = & -3H{A}_{ij} 
  -\frac{1}{a^2e^{2\psi}}\left[R_{ij}[e^{2\psi} \gamma]\right]^{{\rm TF}}
  +\calO(\epsilon^4),  \label{eqn:-evol-K1}\\
 {3\over \alpha} \pa_t H & = &-3H^2-{1\over 2}(E+3P) + \calO(\epsilon^4).
  \label{eqn:evol-K2}
\end{eqnarray}
Note that with the help of the background equations
Eq.~(\ref{eqn:evol-K2}) can be recast into
\begin{align}
\delta P+{ \delta E\over 3}+(\rho_0+P_0)\delta \alpha=\calO(\epsilon^4).
\label{eqn:evol-K3}
\end{align}

The components of the energy-momentum tensor are expanded as
\begin{align}
E&=2XW_X-W+6HG_X\pa_\perp\phi-2XG_\phi+\calO(\epsilon^6),
\label{eqn:E-grad}
\\
P&=W-\pa_\perp G\pa_\perp \phi +\calO(\epsilon^6),
\label{eqn:P-grad}\\
-J_i&=\calJ_0\pa_i (\delta\phi) -(G_X \dot{\phi})_0\pa_i (\delta X)+ 
\calO(\epsilon^5),
\end{align}
where
\begin{align}
&X=(\pa_\perp \phi)^2/2+\calO(\epsilon^6),\\
&\delta X := 
X-X_0=\dot{\phi}_0 \pa_t(\delta \phi) -2 X_0 \delta \alpha
+\calO(\epsilon^4).
\label{eqn:delta-X}
\end{align}
Finally,
noting that $\Box\phi=-\pa_\perp^2\phi-3H\pa_\perp \phi+\calO(\epsilon^4)$, 
the scalar-field equation of motion~(\ref{eqn:EOM-phi}) reduces to
%
\begin{eqnarray}
 W_X(\pa_\perp^2\phi+3H\pa_\perp\phi)+2XW_{XX}\pa_\perp^2\phi+2W_{\phi X} X-W_\phi-2(G_\phi-G_{\phi X}X)(\pa_\perp^2\phi+3H\pa_\perp \phi) \notag\\
+6G_X[\pa_\perp (HX)+3H^2 X]-4XG_{\phi X} \pa_\perp^2\phi-2G_{\phi\phi} X
+6HG_{XX} X\pa_\perp X=\calO(\epsilon^4).
\end{eqnarray}
This equation can also be written in a slightly simpler form as 
\begin{eqnarray}
 \pa_\perp \calJ+3H\calJ =W_\phi-2 XG_{\phi\phi}-2 X G_{\phi X}\pa_\perp^2 \phi
+\calO(\epsilon^4),
  \label{eqn:EOM-phi-grad}
\end{eqnarray}
where 
\begin{align}
\calJ=W_{X}\pa_\perp{\phi}-2G_{\phi}\pa_\perp\phi+6H G_{X} X.
\end{align}
It can be seen that Eq.~(\ref{eqn:EOM-phi-grad}) takes {\em
exactly the same form} as the background scalar-field 
equation of motion~(\ref{eqn:background-phi-J}) under
the identification $\pa_t\leftrightarrow\partial_\perp $.
Now Eq.~(\ref{eqn:E-grad}) can be written using $\calJ$ as 
\begin{align}
E=\calJ \pa_\perp \phi-W+2 X G_\phi.
\label{eqn:E-J-grad}
\end{align}
From Eqs.~(\ref{eqn:EOM-phi-grad}) and  (\ref{eqn:E-J-grad}) we find 
\begin{align}
\pa_\perp E=-3 H(E+P) +\calO(\epsilon^4). 
\label{eqn:conservation}
\end{align}
This equation
is nothing but the conservation law, $n_\mu\na_\nu T^{\mu\nu}=0$.  

Combining Eq.~(\ref{eqn:evol-K2}) with Eq.~(\ref{eqn:conservation}), 
we obtain
%
\begin{equation}
 \partial_t\left[a^2(\delta E)\right]  = O(\epsilon^4).
  \label{eqn:eq-for-delta}
\end{equation}
We can expand Eq.~(\ref{eqn:E-grad}) in 
terms of $\delta \phi$ and $\delta X$, and thus $\delta E$ can be expressed as 
\begin{align}
\delta E=\cale_\phi(t) \delta\phi+\cale_X(t)\delta X+\calO(\epsilon^4).
\label{eqn:exp-E}
\end{align}
This equation relates $\delta\phi$ and $\delta X$ with
a solution to the simple equation~(\ref{eqn:eq-for-delta}).
With the help of Eq.~(\ref{eqn:delta-X}),
Eq.~(\ref{eqn:exp-E}) can be regarded as an equation relating
$\delta\phi$ and $\delta \alpha$.
Similarly, 
one can express $\delta P$ as
\begin{align}
\delta P={1\over a^3}\pa_t\left\{a^3 \left[\calJ_0 (\delta \phi)-(G_X\dot{\phi})_0 (\delta X)
\right]\right\}-(\rho_0+P_0)(\delta \alpha)+\calO(\epsilon^4).
\end{align}
Using Eq.~(\ref{eqn:evol-K3}), one has
\begin{align}
\pa_t\left\{a^3\left[\calJ_0(\delta \phi)-(G_X\dot{\phi})_0 (\delta X)\right]\right\}=
-{a^3\over 3} \delta E+\calO(\epsilon^4),
\label{eqn:exp-P}
\end{align}
which can easily be integrated once to
give another {\em independent} equation
relating $\delta\phi$ and $\delta\alpha$.
In the next section, we will give a general solution
to the above set of equations.

\section{General solution}
\label{sec:solution}

Having thus derived all the relevant equations up to second order in gradient expansion,
let us now present a general solution.  
First, since $\psi=\calO(1)$ and $\partial_t\psi=\calO(\epsilon^2)$, we find
%
\begin{equation}
 \psi = {}^{(0)}C^\psi(\mathbf{x}) + \calO(\epsilon^2),
  \label{eqn:psi-leading}
\end{equation}
where ${}^{(0)}C^\psi(\mathbf{x})$ is an integration constant
which is an arbitrary function of the spatial coordinates $\mathbf{x}$.
Here and hereafter, the superscript $(n)$ indicates that 
the quantity is of order $\epsilon^n$. Similarly, it follows from
${\gamma}_{ij}=\calO(1)$ and
$\partial_t{\gamma}_{ij}=\calO(\epsilon^2)$ that
%
\begin{equation}
 {\gamma}_{ij} = {}{}^{(0)}C^\gamma_{ij}(\mathbf{x}) + \calO(\epsilon^2),
  \label{eqn:gammatilde-leading}
\end{equation}
where ${}^{(0)}C^\gamma_{ij}(\mathbf{x})$ is a $3\times 3$ matrix with a unit
determinant whose components depend only on the spatial
coordinates.

The evolution equations~(\ref{eqn:evol-gamma}) can 
then be integrated to determine the ${\cal O}(\epsilon^2)$ terms in
$\psi$ and $\gamma_{ij}$ as
%
\begin{align}
&\psi={}^{(0)}C^\psi(\mathbf{x})+\int_{t_*}^t H(t') \delta \alpha(t', \mathbf{x}) dt'
 + \calO(\epsilon^4),
\label{eqn:psi-alpha}\\
&\gamma_{ij}={}^{(0)}C^\gamma_{ij}(\mathbf{x})+2 \int_{t_*}^t A_{ij}(t', \mathbf{x})dt'
 + \calO(\epsilon^4),
\label{eqn:gamma-A}
\end{align}
where $t_*$ is some initial time and
integration constants of ${\cal O}(\epsilon^2)$ have been absorbed to
${}^{(0)}C^\psi(\mathbf{x})$ and ${}^{(0)}C_{ij}^\gamma(\mathbf{x})$.

Now Eq.~(\ref{eqn:-evol-K1}) can be integrated to give
\begin{eqnarray}
 A_{ij}={1\over a^3(t)} \left[{}^{(2)}F_{ij}(\mathbf{x})
 \int_{t_*}^t a(t') dt' +{}^{(2)}C^A_{ij}(\mathbf{x})\right]+\calO(\epsilon^4),
\label{eqn: solution-Aij}
\end{eqnarray}
where
\begin{eqnarray}
 {}^{(2)}F_{ij}(\mathbf{x})  &:=&  -\frac{1}{e^{2\psi}}\left[R_{ij}[e^{2\psi} \gamma]\right]^{{\rm TF}}
 \nonumber\\
& = &-\frac{1}{e^{2{}^{(0)}C^\psi}}
  \left[ \left({}^{(2)}{R}_{ij}-\frac{1}{3}{}^{(2)}{R}{}^{(0)}
C^\gamma_{ij}\right)
   + \left(\partial_i{}^{(0)}C^\psi \partial_j{}^{(0)}C^\psi 
   - {}^{(0)}{D}_i {}^{(0)}{D}_j{}^{(0)}C^\psi \right)\right.\nonumber\\
 & & \qquad
 \left.
   - \frac{1}{3}{}^{(0)}C^{\gamma \,kl}
   \left(\partial_k {}^{(0)}C^\psi \partial_l{}^{(0)}C^\psi 
   - {}^{(0)}{D}_k {}^{(0)}{D}_l{}^{(0)}C^\psi  \right) {}^{(0)}C^\gamma_{ij}
  \right].
 \label{eqn:def-F2ij}
\end{eqnarray}
Here,
$^{(0)}C^{\gamma\, kl}$ is the inverse matrix of ${}^{(0)}C^\gamma_{ij}$,
${}^{(2)}{R}_{ij}(\mathbf{x}):=R_{ij}[{}^{(0)}C^\gamma]$
and ${}^{(2)}{R}(\mathbf{x}):=R[{}^{(0)}C^\gamma]$
are the Ricci tensor and the Ricci scalar
constructed from the zeroth-order spatial metric $^{(0)}C_{ij}^\gamma(\mathbf{x})$,
and ${}^{(0)}{D}$ is the covariant derivative
associated with ${}^{(0)}C^\gamma_{ij}$. 
Note that $^{(0)}C^{\gamma\, ij}\,{}^{(2)}F_{ij}=0$
by definition. 
The integration constant, ${}^{(2)}C^A_{ij}(\mathbf{x})$,
is a symmetric matrix whose components depend 
only on the spatial coordinates and which satisfies the traceless condition 
$^{(0)}C^{\gamma\, ij}\,{}^{(2)}C^A_{ij}=0$.
Substituting the above result to Eq.~(\ref{eqn:gamma-A}),
we arrive at
%
\begin{equation}
 {\gamma}_{ij} = {}^{(0)}C^\gamma_{ij}(\mathbf{x}) + 2\left[
      {}^{(2)}F_{ij}(\mathbf{x})\int^t_{t_*}\frac{dt'}{a^3(t')}\int^{t'}_{t_*}a(t'')dt''
      + {}^{(2)}C^A_{ij}(\mathbf{x})\int^t_{t_*}\frac{dt'}{a^3(t')}\right]
  + \calO(\epsilon^4).
  \label{eqn:gammatilde-sol}
\end{equation}

Next, it is straightforward to integrate Eq.~(\ref{eqn:eq-for-delta})
to obtain
%
\begin{equation}
 \delta E = 
\frac{1}{a^2(t)}{}^{(2)}{\cal K}(\mathbf{x}) + \calO(\epsilon^4),
 \label{eqn: solution-delta}
\end{equation}
where ${}^{(2)}{\cal K}(\mathbf{x})$ is an arbitrary
function of the spatial coordinates.  
With this solution for $\delta E$, Eqs.~(\ref{eqn:exp-E}) and~(\ref{eqn:exp-P})
reduce to
\begin{align}
&\cale_\phi(t) \delta\phi+\cale_X(t)\delta X= 
\frac{1}{a^2(t)}{}^{(2)}{\cal K}(\mathbf{x}) +\calO(\epsilon^4),
\label{eqn:exp-E2}\\
&\pa_t\left\{a^3\left[\calJ_0(\delta \phi)-(G_X\dot{\phi})_0 (\delta X)\right]\right\}=
-{{1} \over 3 a^2(t)} {}^{(2)}{\cal K}(\mathbf{x})+\calO(\epsilon^4),
\label{eqn:exp-P1}
\end{align}
and the latter equation can further be integrated to give 
\begin{align}
\calJ_0(\delta \phi)-(G_X\dot{\phi})_0 (\delta X)=
{{}^{(2)}C^\chi(\mathbf{x}) \over a^3(t)}-{{}^{(2)}{\cal K}(\mathbf{x}) \over 3 a^3(t)} \int_{t_*}^t a(t') dt'+\calO(\epsilon^4),
\label{eqn:exp-P2}
\end{align}
where we have introduced
another integration constant $^{(2)}C^\chi (\mathbf{x})$.

With the help of Eqs.~(\ref{eqn:delta-X}), one can solve the system of equations~(\ref{eqn:exp-E2})
and~(\ref{eqn:exp-P2}), leading to 
\begin{align}
\delta \phi={1\over \cal A}\left\{\left[(G_X\dot{\phi})_0
-{\cale_X(t) \over 3 a(t)}\int_{t_*}^t a(t') dt' \right]
\frac{{}^{(2)}{\cal K}}{a^2(t)}+{\cale_X(t)\over a^3(t)} {}^{(2)}C^\chi\right\}+\calO(\epsilon^4),
\end{align}
and 
\begin{align}
\delta\alpha=\pa_t\left({\delta \phi\over \dot{\phi}_0}\right)
-{1\over 2 X_0 \calg (t)} \left\{
\left[
1-{\Theta(t)\over a(t)}\int_{t_*}^t a(t') dt' \right]
\frac{{}^{(2)}{\cal K}}{a^2(t)}
+{3\Theta(t)\over a^3(t)} {}^{(2)}C^\chi\right\}+\calO(\epsilon^4),
\label{solalpha}
\end{align}
where ${\calA}:=(\cale_\phi G_X \dot{\phi}+\cale_X \calJ)|_{0}$.
In deriving the above solution
we have used the background scalar-field equation of motion~(\ref{eqn:background-gphi}).

Finally, substituting Eq.~(\ref{solalpha}) to Eq.~(\ref{eqn:psi-alpha}), we obtain
\begin{align}
\psi&= {}^{(0)}C^\psi
+\int_{t_*}^t dt' H(t') \pa_{t'} \left(
{\delta \phi\over \dot{\phi}_0}\right) -\int_{t_*}^t dt'
{H(t')\over 2 X_0 \calg (t')} \left\{\left[1-{\Theta(t')\over a(t')}
\int_{t_*}^{t'} a(t'') dt'' \right]\frac{{}^{(2)}{\cal K}}{a^2(t')}
+{3\Theta(t')\over a^3(t')} {}^{(2)}C^\chi\right\}+{\cal O}(\epsilon^4)
\notag\\&= {}^{(0)}C^\psi(\mathbf{x})+ {H \delta \phi\over \dot{\phi}_0}+
\int_{t_*}^t dt' {(\rho_0+P_0) \delta\phi \over 2 \dot{\phi}_0}
\notag\\
&\qquad
-\int_{t_*}^t dt'{H (t')\over 2 X_0 \calg a^2\, (t') } 
\left\{  \left[1-{\Theta(t')\over a(t')}\int_{t_*}^{t'} a(t'') dt'' \right]
{}^{(2)}{\cal K}+{3\Theta(t')\over a(t')} {}^{(2)}C^\chi\right\}
+\calO(\epsilon^4),
\end{align}
where we performed integration by parts and used the background equation. 

So far we have introduced five integration constants,
${}^{(0)}C^\psi(\mathbf{x})$, ${}^{(0)}C^\gamma_{ij}(\mathbf{x})$,
${}^{(2)}C^A_{ij}(\mathbf{x})$, ${}^{(2)}{\cal K}(\mathbf{x})$, and ${}^{(2)}C^\chi(\mathbf{x})$,
upon solving the field equations up to ${\cal O}(\epsilon^2)$.
Here, it should be pointed out that they are not
independent.
Indeed, 
Eqs.~(\ref{eqn:Hamiltonian}) and~(\ref{eqn:Momentum})
impose the following constraints among the integration constants:
%
\begin{eqnarray}
 {}^{(2)}{\cal K}(\mathbf{x}) & =&
 \frac{{}^{(2)}\hat R(\mathbf{x})}{2}
   + \calO(\epsilon^4), 
  \nonumber\\
  e^{-3 {}^{(0)} C^\psi} {}^{(0)}C^{\gamma\, j k} {}^{(0)}D_j
  \left[e^{3 {}^{(0)}C^\psi} \,{}^{(2)}C^A_{ki}(\mathbf{x})\right]
  & =&  \pa_i\, {}^{(2)}C^\chi(\mathbf{x}) + \calO(\epsilon^5),
  \notag\\
 e^{-3 {}^{(0)} C^\psi} {}^{(0)}C^{\gamma\, j k} {}^{(0)}D_j\left[
 e^{3 {}^{(0)}C^\psi} \,{}^{(2)}F_{ki}(\mathbf{x})\right]
 & =& -{1\over 6} \pa_i\, {}^{(2)}\hat R(\mathbf{x}) + \calO(\epsilon^5), 
   \label{eqn: constraint-integration-const}
\end{eqnarray}
where
$^{(2)}\hat R(\mathbf{x}):=R[e^{2 {}^{(0)}C^\psi}\, {}^{(0)}C^\gamma]$ is
the Ricci scalar constracted from the metric $e^{2{}^{(0)}C^\psi} \,{}^{(0)}C^\gamma_{ij}$. Here, $^{(2)}\hat R(\mathbf{x})$ should not be confused with $^{(2)}R(\mathbf{x})$.
The latter is the Ricci scalar constructed from $^{(0)}C_{ij}^\gamma$
and not from $e^{2{}^{(0)}C^\psi} \,{}^{(0)}C^\gamma_{ij}$.
Explicitly,
\begin{align}
{}^{(2)}\hat R(\mathbf{x})= \left[{}^{(2)}R(\mathbf{x})
-2 \left(2 {}^{(0)}D^2 {}^{(0)}C^\psi +{}^{(0)}C^{\gamma\, ij}
\partial_i {}^{(0)}C^\psi \partial_j {}^{(0)}C^\psi \right)\right]e^{-2 {}^{(0)}C^\psi}.
\label{def:K22}
\end{align}
Note that the third equation is automatically 
satisfied provided that the last equation holds, 
as can be verified by using Eq.~(\ref{eqn:def-F2ij}).


In summary, we have
integrated the field equations up to second order in gradient expansion
and obtained the following solution for generic single-field inflation:
%
\begin{eqnarray}
\delta E & =  & \frac{{}^{(2)} \hat R(\mathbf{x})}{2a^2}
   + \calO(\epsilon^4), \nonumber\\
 \delta \phi &=&{1\over {\cal A} a^2}\left[\left((G_X\dot{\phi})_0
-{\cale_X \over 3 a}\int_{t_*}^t a(t') dt' \right){{}^{(2)}\hat R(\mathbf{x})\over 2}
+{\cale_X\over a} {}^{(2)}C^\chi(\mathbf{x})\right]+\calO(\epsilon^4), \nonumber\\
\delta \alpha  &=  & \pa_t\left({\delta \phi\over \dot{\phi}_0}\right)
-{1\over 2 X_0 \calg  a^2} \left[ \left(1-{\Theta\over a}\int_{t_*}^t a(t') dt' \right)
{{}^{(2)}\hat R(\mathbf{x})\over 2}
+{3\Theta\over a} {}^{(2)}C^\chi(\mathbf{x})\right]+\calO(\epsilon^4), \nonumber\\
 \psi  &= & {}^{(0)}C^\psi(\mathbf{x})
 + {H \delta \phi\over \dot{\phi}_0}+
\int_{t_*}^t dt' {(\rho_0+P_0) \delta\phi \over 2 \dot{\phi}_0}\notag\\
&&-\int_{t_*}^t dt'{H \over 2 X_0 \calg a^2 } 
\left[  \left(1-{\Theta \over a}\int_{t_*}^{t'} a(t'') dt'' \right)
{{}^{(2)}\hat R(\mathbf{x})\over 2}+{3\Theta\over a} {}^{(2)}C^\chi(\mathbf{x})\right]
+\calO(\epsilon^4), \nonumber\\
{A}_{ij}  &= & \frac{1}{a^3}
  \left[{}^{(2)}F_{ij}(\mathbf{x})\int_{t_0}^t a(t')dt' + {}^{(2)}C^A_{ij}(\mathbf{x})\right] 
  + O(\epsilon^4), \notag\\
 {\gamma}_{ij} & =  & {}^{(0)}C^\gamma_{ij}(\mathbf{x})+2\left[
      {}^{(2)}F_{ij}(\mathbf{x})
	  \int^t_{t_*}\frac{dt'}{a^3(t')}\int^{t'}_{t_*}a(t'')dt''   + {}^{(2)}C^A_{ij}(\mathbf{x})
	  \int^t_{t_*}\frac{dt'}{a^3(t')}\right]
  + \calO(\epsilon^4). \label{eqn:general solution}
\end{eqnarray}
The $\mathbf{x}$-dependent integration constants,
${}^{(0)}C^\psi$, ${}^{(0)}C^\gamma_{ij}$, ${}^{(2)}C^\chi$  and ${}^{(2)}C^A_{ij}$,
satisfy the following conditions:
%
\begin{eqnarray}
 &&{}^{(0)}C^\gamma_{ij}=  {}^{(0)}C^\gamma_{ji}, \quad 
\det({}^{(0)}C^\gamma_{ij}) = 1,\quad
{}^{(2)}C^A_{ij}  =  {}^{(2)}C^A_{ji}, 
\quad {}^{(0)}C^{\gamma\,ij} \, {}^{(2)}C^A_{ij} = 0,
  \nonumber\\&&
  e^{-3 ^{(0)} C^\psi}
  {}^{(0)}C^{\gamma\, j k} {}^{(0)}D_j\left(e^{3 ^{(0)}C^\psi} \,{}^{(2)}C^A_{ki}\right)
  =   \pa_i\, {}^{(2)}C^\chi. 
\end{eqnarray}

Before closing this section,
we remark that the gauge condition~(\ref{time-orthogonal-cond}) remains
unchanged under a purely spatial coordinate transformation
%
\begin{equation}
 x^i \to \bar x^i = f^i(\mathbf{x}).
\end{equation}
This means that the zeroth-order spatial metric ${}^{(0)}C^\gamma_{ij}$ 
contains three residual gauge degrees of freedom.
Therefore, the number of degrees of freedom associated
with each integration constant is summarized as follows: 
%
\begin{eqnarray}
 {}^{(0)}C^\psi & \cdots & 
  1 \mbox{ scalar growing mode}
  = 1 \mbox{ component}, \nonumber\\
 {}^{(0)}C^\gamma_{ij} & \cdots & 
  2 \mbox{ tensor growing modes}
  = 5 \mbox{ components} - 3 \mbox{ gauge}, \nonumber\\
 {}^{(2)}C^\chi & \cdots & 
  1 \mbox{ scalar decaying mode}
  = 1 \mbox{ component}, \nonumber\\
 {}^{(2)}C^A_{ij} & \cdots & 
  2 \mbox{ tensor decaying modes}
  = 5 \mbox{ components} - 3 \mbox{ constraints}. 
\end{eqnarray}

\section{Nonlinear curvature perturbation}
\label{sec:nlcurvpert}

In this section, we will define a new variable which is a nonlinear
generalization of the curvature perturbation up to
$\calO(\epsilon^2)$ in gradient expansion.
We will show that this variable satisfies a nonlinear 
second-order differential equation,
and, as in Ref.~\cite{Takamizu:2010xy},
the equation can be deduced
as a generalization of
the corresponding linear perturbation equation.
To do so, one should notice the following fact
on the definition of the {\em curvature perturbation}:
in linear theory the curvature perturbation is named so because it
is directly related to
the three-dimensional Ricci scalar;
$\psi$ may be called so at fully nonlinear order in perturbations
and at leading order in gradient expansion;
and, as pointed out in Ref.~\cite{Takamizu:2010xy},
$\psi$ is no longer appropriate to be called so at second order in gradient expansion.
To define the curvature perturbation appropriately at ${\cal O}(\epsilon^2)$,
one needs to take into account the contribution from $\gamma_{ij}$.
Let us denote this contribution as $\chi$.
We carefully define the curvature perturbation as a sum of $\psi$ and $\chi$
so that the new variable reproduces the correct result in the linear limit.
In what follows we remove the subscript $0$ 
from the background quantities since there will be no danger of confusion.

\subsection{Assumptions and definitions}
\label{subsec:assumption}

As mentioned in the previous section,
we still have residual spatial gauge degrees freedom, which we are going to
fix appropriately.
To do so, we assume that the contribution from gravitational waves
to $\gamma_{ij}$ is negligible
and consider
the contribution from scalar-type perturbations only.
We may then choose the spatial coordinates so that
$\gamma_{ij}$ coincides with
the flat metric
at sufficiently late times during inflation,
\begin{eqnarray}
{\gamma}_{ij}\to\delta_{ij}\quad (t\to\infty).
\label{eq: gamma-ij t-infty}
\end{eqnarray}
In reality, the limit $t\to\infty$ may be reasonably interpreted
as $t\to t_{\rm late}$
where $t_{\rm late}$ is some time close to the end of inflation.
Up to ${\cal O}(\epsilon^2)$,
this condition completely removes the residual three gauge degrees of freedom.

We wish to define
appropriately a nonlinear curvature perturbation on uniform $\phi$ hypersurfaces
($\delta \phi(t,\mathbf{x})=0$)
and derive a nonlinear evolution equation for the perturbation.
The nonlinear result on uniform $\phi$ hypersurfaces is
to be compared with
the linear result for G-inflation~\cite{Kobayashi:2010cm}. 
However, in the previous section the general solution
was derived in the uniform expansion gauge.
For our purpose we will therefore go
from the uniform expansion gauge to the uniform $\phi$ gauge\footnote{The gauge
in which $\phi$ is uniform is sometimes called the unitary gauge.
The unitary gauge does not coincide with the comoving gauge in G-inflation,
as emphasized in~\cite{Kobayashi:2010cm}.}.
It is clear that
at leading order in gradient expansion
the uniform expansion gauge
coincides with the uniform $\phi$ gauge. 
In this case, it would be appropriate simply to define $\psi$ to be
the nonlinear curvature perturbation.
At second order in gradient expansion, however,
this is not the correct way of defining the nonlinear curvature perturbation.
We must extract the appropriate scalar part $\chi$ from $\gamma_{ij}$,
which will yield an extra contribution to the total curvature perturabtion,
giving a correct definition of the nonlinear
curvature perturbation at ${\cal O}(\epsilon^2)$.

Let us use the subscripts $K$ and $u$ to indicate
the quantity in the uniform $K$ and $\phi$ gauges, respectively,
so that in what follows the subscript $K$ is attached to the solution
derived in the previous section.
First, we derive the relation between $\psi_K$ and $\psi_u$ up to 
$\calO(\epsilon^2)$. 
In general, one must consider a nonlinear transformation
between different time slices. 
The detailed description on this issue can be found 
in Ref.~\cite{Naruko:2012fe}. 
However, thanks to the fact that $\delta \phi_K=\calO(\epsilon^2)$,
one can go from the uniform $K$ gauge to the uniform $\phi$ gauge
by the transformation analogous to
the familiar linear gauge transformation.
Thus, $\psi_u$ is obtained as
\begin{eqnarray}
{\psi}_u={\psi}_K-\frac{H}{\dot{\phi}}\delta \phi_K+\calO(\epsilon^3). 
\label{def: comoving nonlinear curvature perturb}
\end{eqnarray}
One might think that the shift vector $\beta^i_u$ appears
in this new variable
as a result of the gauge transformation, but $\beta^i$ can always be gauged away
by using a spatial coordinate transformation.
The general solution for ${\psi}_u$ valid up to 
$\calO(\epsilon^2)$
is thus given by the linear combination
of the solution for $\psi_K$ and $\delta \phi_K$ displayed in 
Eq.~(\ref{eqn:general solution}).
Note here that the spatial metric $\gamma_{ij}$
remains the same at $\calO(\epsilon^2)$ accuracy under the change
from the uniform $K$ gauge to the uniform $\phi$ gauge:
\begin{eqnarray}
\gamma_{ij\, K}=\gamma_{ij\,u}+\calO(\epsilon^4)\,.
\end{eqnarray}

We now turn to the issue of appropriately defining a nonlinear
curvature perturbation to $\calO(\epsilon^2)$ accuracy.
Let us denote the linear curvature perturbation in
the uniform $\phi$ gauge by ${\cal R}^{\rm Lin}_u$. 
%
%
In the linear limit, $\psi$ reduces to the longitudinal component
$H^{\rm Lin}_L$
of scalar perturbations, while 
$\chi$ to
traceless component $H^{\rm Lin}_T$: 
\begin{align}
\psi\to H^{\rm Lin}_L, \quad
\chi\to H^{\rm Lin}_T.
\end{align}
The linear curvature perturbation is given by 
${\cal R}^{\rm Lin}=(H^{\rm Lin}_L+H^{\rm Lin}_T/3)Y$.
Here, 
we have followed Ref.~\cite{Kodama:1985bj}
and the perturbations are expanded in scalar harmonics $Y$
satisfying $\left(\partial_i\partial^i+k^2\right)Y_{\mathbf{k}}=0$, with the summation over
$\mathbf{k}$ suppressed for simplicity. 
The spatial metric in the linear limit is expressed as 
\begin{eqnarray}
\hat\gamma_{ij}^{{\rm Lin}}=a^2\left(\delta_{ij}
+2 H^{\rm Lin}_L Y \delta_{ij}+2 H^{\rm Lin}_T Y_{ij}\right)\,,
\end{eqnarray}
where
$Y_{ij}=k^{-2} \left[\partial_i \partial_j -(1/ 3) \delta_{ij} 
\partial_l\partial^l \right] Y_{\mathbf{k}}$.
Since $\psi$ corresponds to $H^{\rm Lin}_L$, one can read off from
the above expression that
${\gamma}_{ij} = \delta_{ij} + 2H^{\rm Lin}_T Y_{ij}$ in the linear limit. 
Thus, our task is to extract from $\gamma_{ij}$
the scalar component $\chi$ that 
reduces to $H^{\rm Lin}_T $ in linear limit. 
It was shown in Ref.~\cite{Takamizu:2010xy} 
that by using
the inverse Laplacian operator on the flat background, $\Delta^{-1}$,
one can naturally define $\chi$ as
\begin{eqnarray}
\chi :=
-\frac{3}{4}\Delta^{-1}\left[\partial^i e^{-3\psi}
\partial^j e^{3\psi}({\gamma}_{ij}-\delta_{ij}) \right].
\label{def: nonlinear HT0}
\end{eqnarray}
In terms of $\chi$ defined above,
the nonlinear 
curvature perturbation is defined, to $\calO(\epsilon^2)$, as
\begin{eqnarray}
{\frR}^{\rm NL}:= {\psi}\,+\,{\chi\over 3}\,.
\label{def0: nonlinear variable zeta}
\end{eqnarray}

As is clear from Eq.~(\ref{def: nonlinear HT0}), extracting $\chi$ generally
requires a spatially nonlocal operation.
However, as we will see
in the next subsection, in the uniform $\phi$ gauge 
supplemented with the asymptotic condition on the spatial 
coordinates~(\ref{eq: gamma-ij t-infty}), it
is possible to obtain the explicit expression for the nonlinear
version of $\chi$ from our solution~(\ref{eqn:general solution}) 
without any nonlocal operation. 

%
%
%
%
\subsection{Solution}
\label{subsec:explicit}

We start with presenting an explicit expression for $\psi_u$.
It follows from
Eqs.~(\ref{eqn:general solution}) and~(\ref{def: comoving nonlinear curvature perturb})
that
\begin{eqnarray}
{\psi}_u={}^{(0)}C^\psi(\mathbf{x})+{}^{(2)}C^\psi(\mathbf{x})
+{f}_R(t)\,{}^{(2)}\hat R(\mathbf{x})+
{f}_\chi (t)\, {}^{(2)}C^\chi(\mathbf{x}) +\calO(\epsilon^4).
\label{sol: general sol R}
\end{eqnarray}
Note here that although the integration constant
${}^{(2)}C^\psi(\mathbf{x})$
was absorbed into the redefinition of ${}^{(0)}C^\psi(\mathbf{x})$
in the previous section,
we do not do so
in this section for later convenience. 
The time-dependent functions $f_R(t)$ and $f_C(t)$ are defined as
\begin{eqnarray}
{f}_R(t)&:=&
\int_{t_*}^t {dt' \over 2 a^2}  \left\{{(\rho+P)\over 
2 \dot{\phi} \calA } 
\left[ G_X\dot{\phi}-{\cale_X\over 3 a}
\int_{t_*}^{t'} a(t'') dt'' \right]-{H\over 2 X \calg } 
\left( 1-{\Theta \over a}\int_{t_*}^{t'} a(t'') dt'' \right)\right\},
\label{eq: f-K t} \\
{f}_\chi(t)
&:=&\int_{t_*}^t {dt' \over a^3}  \left[{(\rho+P) \cale_X \over 
2 \dot{\phi} \calA } -{3 H \Theta \over 2 X\calg }\right].
\label{eq: f-C t} 
\end{eqnarray}

Since $\gamma_{ij\,u}$ coincides with $\gamma_{ij\, K}$ up to ${\cal O}(\epsilon^2)$,
it is straightforward to see
\begin{eqnarray}
{\gamma}_{ij\,u}={}^{(0)}C^\gamma_{ij}(\mathbf{x})+{}^{(2)}C^\gamma_{ij}(\mathbf{x})+
2g_F(t){}^{(2)}F_{ij}(\mathbf{x})
+2 g_A(t) {}^{(2)}C^A_{ij}(\mathbf{x})+\calO(\epsilon^4),
\label{eq: til-gamma-ij}
\end{eqnarray}
where
\begin{eqnarray}
g_F(t):=\int_{t_*}^{t}
{dt'\over a^3(t')}\int_{t_*}^{t'} a(t'') dt''\,, 
\quad
g_A(t):=\int_{t_*}^t {dt'\over a^3(t')}\,. 
\label{def: integral-A-B}
\end{eqnarray}
The integration constants ${}^{(0)}C^\gamma_{ij}$ and ${}^{(2)}C^\gamma_{ij}$
are determined from the condition~(\ref{eq: gamma-ij t-infty}) as
\begin{eqnarray}
&& {}^{(0)}C^\gamma_{ij} = \delta_{ij}\,,\nonumber\\
&& {}^{(2)}C^\gamma_{ij} =
-2 g_F(\infty){}^{(2)}F_{ij}-2g_A(\infty) {}^{(2)}C^A_{ij}.
\label{def: f2-ij}
\end{eqnarray}
We now have ${}^{(0)}C_{ij}^\gamma =\delta_{ij}$,
and hence ${}^{(2)}R_{ij}(\mathbf{x})=R_{ij}[{}^{(0)}C^\gamma]=0$.
This simplifies the explicit expression for ${}^{(2)}\hat R(\mathbf{x})$
and ${}^{(2)}F_{ij}(\mathbf{x})$; they are given solely in terms of ${}^{(0)}C^\psi$
and the usual derivative operator $\partial_i$.

Substituting Eq.~(\ref{eq: til-gamma-ij}) to 
the definition~(\ref{def: nonlinear HT0}),
we obtain
\begin{eqnarray}
{\chi_u\over 3}=
{{}^{(2)}\hat R(\mathbf{x})\over 12}\left[g_F(t)-g_F(\infty)\right]
-{{}^{(2)}C^\chi(\mathbf{x})\over 2}\left[g_A(t)-g_A(\infty)\right]
+\calO(\epsilon^4).
\label{eq: HT}
\end{eqnarray}
It is easy to verify that the linear limit of $\chi_u$
reduces consistently to $H_T^{\rm Lin}Y$. 
We then finally arrive at the following
explicit solution for the appropriately defined nonlinear curvature perturbation
in the uniform $\phi$ gauge:
\begin{eqnarray}
{\frR}_u^{\rm NL}
= {}^{(0)}C^\psi(\mathbf{x})+{}^{(2)}C^\psi(\mathbf{x})
+{}^{(2)}\hat R(\mathbf{x})\left[ f_R(t)+
\frac{g_F(t)}{12}-\frac{g_F(\infty)}{12}
\right]
+{}^{(2)}C^\chi(\mathbf{x})\left[ f_\chi(t)
-\frac{g_A(t)}{2}+\frac{g_A(\infty)}{2}\right].
\label{def: nonlinear variable zeta}
\end{eqnarray}

Let us comment on the dependence of $\frR_u^{\rm NL}$
on the initial fiducial time $t_*$.
One may take $t_*$ as the time when
our nonlinear superhorizon solution is matched
to the perturbative solution whose initial condition
is fixed deep inside the horizon.
Then, $\frR_u^{\rm NL}$ should not depend on the choice of $t_*$, though
apparent dependences are found in the lower bounds of the integrals
${f}_R(t)$, ${f}_\chi(t)$, $g_F(t)$, and $g_A(t)$.
Actually,
in the same way as discussed in Ref.~\cite{Takamizu:2010xy}, 
one can check that $\frR_u^{\rm NL}$ is 
invariant under the infinitesimal shift $t_*\to t_*+\delta t_*$.
%
%
\subsection{Second-order differential equation}
\label{subsec:nleq}
%
%
Having obtained explicitly the solution
$\frR_u^{\rm NL}$ in Eq.~(\ref{def: nonlinear variable zeta}),
now we are going to deduce the second-order differential equation
that $\frR_u^{\rm NL}$ obeys at ${\cal O}(\epsilon^2)$ accuracy.
For this purpose, we rewrite
$f_R(t)$ and $f_\chi(t)$ in terms of
\begin{eqnarray}
z:={a\dot{\phi}\sqrt{\calg} \over \Theta}\,.
\label{def: variable-z}
\end{eqnarray}
This is a generalization of familiar ``$z$''
in the Mukhanov-Sasaki equation~\cite{Mukhanov:1990me},
and reduces indeed to $a\sqrt{(\rho+P)}/H c_s$
in the case of k-inflation. 
With some manipulation, it is found that $f_R(t)$ and $f_\chi(t)$ can be rewritten as
\begin{eqnarray}
{f}_R(\eta)&=&{1\over 2}\int_{\eta_*}^{\eta}
{a(\eta') d\eta'\over 
z^2 \Theta\,(\eta')}+{1\over 2}\int_{\eta_*}^{\eta}
{d\eta'\over z^2(\eta')}\int_{\eta_*}^{\eta'} a(\eta'') d\eta'' 
-{1\over 12} \int_{\eta_*}^{\eta} {d\eta'\over a^2(\eta')}
\int_{\eta_*}^{\eta'} a^2 (\eta'') d\eta'',\nonumber\\
{f}_\chi(\eta)&=&{1\over 2}\int_{\eta_*}^\eta {d\eta'\over a^2(\eta')} 
-3 \int_{\eta_*}^{\eta}{d\eta'\over z^2(\eta')}\,,
\end{eqnarray}
where the conformal time defined by $d\eta=dt/a(t)$ was used instead of $t$, and
$\eta_*$ corresponds to the fiducial initial time.
Further, it is convenient to express them in the form
\begin{eqnarray}
{f}_R(\eta)=
F(\eta_*)-F(\eta)-{1\over 12} g_F(\eta) \,,\qquad
{f}_\chi(\eta)={1\over 2} g_A(\eta)
+D(\eta_*)-D(\eta),
\end{eqnarray}
where we defined
\begin{eqnarray}
D(\eta)=3 \int_{\eta}^{0}{ d\eta' \over z^2(\eta')}\,, 
\quad
F(\eta)={1 \over 2 }\int_{\eta}^{0} 
{d\eta'\over z^2(\eta')}
\int_{\eta_*}^{\eta'} a^2 (\eta'') d\eta''- {1 \over 2 }
\int_{\eta}^{0} 
{a(\eta')d\eta'\over z^2\Theta (\eta')}\,.
\label{def: integral-D-F}
\end{eqnarray}
The functions $D(\eta)$ and $F(\eta)$
are defined so that $D, F\to 0$ as $\eta\to 0$.
It is important to notice that $D(\eta)$ is
the decaying mode in the long-wavelength limit, {\em i.e.}, at leading order in
gradient expansion, in the linear theory,
satisfying
\begin{eqnarray}
D''+2\frac{z'}{z}D'=0\,,
\end{eqnarray}
while $F(\eta)$ is the ${\cal O}(k^2)$
correction to the growing (constant) mode satisfying
\begin{eqnarray}
F''+2\frac{z'}{z}F'+c_s^2=0\,,
\label{eq:DFmeaning}
\end{eqnarray}
where we assume that the growing mode solution is
of the form $1+k^2F(\eta)+{\cal O}(k^4)$.
In the above equations the prime stands for differentiation
with respect to the conformal time and $c_s^2$ is the sound speed squared
of the scalar fluctuations defined as
\begin{align}
c^2_s:= {{\cal F}(t)\over \calg(t)},\quad
{\cal F}(t):={1\over X_0} (-\pa_t \Theta+\Theta G_X X
\dot{\phi})_0. 
\end{align}

Using $D$ and $F$, Eq.~(\ref{def: nonlinear variable zeta})
can be written as
\begin{eqnarray}
{\frR}_u^{\rm NL}(\eta)
=
{}^{(0)}C^\psi(\mathbf{x})
+{}^{(2)}C^{\frR}(\mathbf{x})
-{}^{(2)}\hat R(\mathbf{x})F(\eta)
-{}^{(2)}C^{\chi}(\mathbf{x})D(\eta)
+\calO(\epsilon^4),
\label{eq: sol-tild-zeta}
\end{eqnarray}
%
%
%
where time-independent terms of ${\cal O}(\epsilon^2)$
are collectively absorbed to ${}^{(2)}C^{\frR}(\mathbf{x})$.
It turns out that the solution can be expressed
simply in terms of the two time-dependent functions
corresponding to the decaying mode and the ${\cal O}(k^2)$
correction to the growing mode in the linear theory.
This shows that, within ${\cal O}(\epsilon^2)$ accuracy in gradient expansion,
the curvature perturbation ${\frR}_u^{\rm NL}$ obeys the following nonlinear
second-order differential equation:
%
%
\begin{eqnarray}
{{\frR}_u^{\rm NL}}''+2 {z'\over z} 
{{\frR}_u^{\rm NL}}' +{c_s^2\over 4} \,{}^{(2)}R[\,
{\frR}_u^{\rm NL}\,]=\calO(\epsilon^4)\,,
\label{eq: basic eq for NL}
\end{eqnarray}
where ${}^{(2)}R[\,{\frR}_u^{\rm NL}\,]$ is the Ricci scalar
of the metric $\delta_{ij}\exp\left(2\frR_u^{\rm NL}\right)$.
Equation~(\ref{eq: basic eq for NL}) is our main result. 
It is easy to see that in the linear limit
Eq.~(\ref{eq: basic eq for NL}) reproduces the previous result
for the curvature perturbation in the unitary gauge~\cite{Kobayashi:2010cm},
\begin{eqnarray}
{{\cal R}^{\rm Lin}_u}''+2{z'\over z} {{\cal R}^{\rm Lin}_u}'
-c_s^2\,\Delta {\cal R}^{\rm Lin}_u=0\,,
\label{eq: linear eq R}
\end{eqnarray}
where $\Delta$ denotes the Laplacian operator on the flat background. 

Equation~(\ref{eq: basic eq for NL}) can be regarded as the master
equation for the nonlinear superhorizon curvature perturbation at
second order in gradient expansion. It must, however, be used with caution,
since it is derived under the assumption that the
decaying mode is negligible at leading order in gradient expansion. Moreover, if one the right-hand side of (\ref{eq: basic eq for NL}) set to exactly
zero, this master equation becomes a closed equation, and it 
be a useful approximation to a full nonlinear solution 
on the Hubble horizon scales or even on scales somewhat smaller than
the Hubble radius.

\section{Summary and discussion}
\label{sec:summary}

In this paper, we have developed a theory of nonlinear cosmological perturbations on
superhorizon scales for G-inflation, for which
the inflaton Lagrangian is given by
$W(X,\phi)-G(X,\phi)\Box\phi$. 
In the case of $G_X=0$, {\em i.e.}, k-inflation,
the energy-momentum tensor for the scalar field is equivalent to
that of a perfect fluid. In the case of G-inflation, however,
it can no longer be recast into a perfect fluid form,
and hence its imperfect nature shows up
when the inhomogeneity of the Universe is considered. 
We have solved the field equations using spatial gradient expansion
in terms of a small parameter $\epsilon:=k/(aH)$, where $k$
is a wavenumber, and obtained
a general solution for the metric and the scalar field up to $\calO(\epsilon^2)$.

We have introduced an appropriately defined variable for 
the nonlinear curvature perturbation in the uniform $\phi$ gauge,
${\frR}_u^{\rm NL}$. Upon linearization, this variable
reduces to the previously defined linear curvature perturbation
${\cal R}_u^{\rm Lin}$ on uniform $\phi$ hypersurfaces.
Then, it has been shown that ${\frR}_u^{\rm NL}$
satisfies a nonlinear second-order differential equation~(\ref{eq: basic eq for NL}),
which is a natural extension of the linear perturbation equation for ${\cal R}_u^{\rm Lin}$.
We believe that our result can further be extended to 
include generalized G-inflation, {\em i.e.},
the most general single-field inflation model~\cite {Kobayashi:2011nu},
though the computation required would be much more complicated.

The nonlinear evolution of perturbations, and hence the 
amount of non-Gaussianity, 
are affected by the $\calO(\epsilon^2)$ corrections if, for example, 
there is a stage during which the slow-roll conditions are violated. 
Calculating the three point correlation function of curvature perturbations 
including the $\calO(\epsilon^2)$ corrections will be addressed in a 
future publication. 
Finally we have comment on our method compared to the in-in formalism developed in the literature. Our formalism is vaid on the classical evolution in superhorizon scales, while the in-in formalism can also calculate a quantum evolution on sub-horizon scale. 
So comparison with each other leads to picking out the quantum effect of non-Gaussianity directly. We have handled the curvature perturbation itself in our formalism, not the correlation function in the in-in one, then its time evolution is more clearly understood.

\acknowledgments
This work was supported by the JSPS Grant-in-Aid for Young Scientists
(B) No.~23740170 and for JSPS Postdocoral Fellowships No. 24-2236.


\end{document}